\documentclass[nonacm,sigplan]{acmart}

\usepackage[compact]{titlesec} 
\usepackage{tikz}
\usepackage{amsmath,pifont} 
\usepackage[english]{babel} 
\usepackage{algorithm}
\usepackage{algorithmicx}
\usepackage[noend]{algpseudocode}

\usepackage[]{hyperref}
\PassOptionsToPackage{hyphens}{url}\usepackage{hyperref}
\hypersetup{colorlinks={true},linkcolor={blue},citecolor=blue}
\usepackage[capitalise]{cleveref}
\crefname{section}{}{\S\S}
\crefdefaultlabelformat{\S#2#1#3}
 
\makeatletter
\def\BState{\State\hskip-\ALG@thistlm}
\renewcommand\theHALG@line{\thealgorithm.\arabic{ALG@line}}
\makeatother

\usepackage{utfsym}
\usepackage{xspace}
\usepackage{comment}
\usepackage[normalem]{ulem}
\usepackage{enumitem}
\setlist[enumerate]{itemsep=0mm}
\usepackage{multirow}
\usepackage{float}
\usepackage{makecell}

\setlist{nolistsep}

\usepackage{listings}
\usepackage{xcolor}
\usepackage[font=small]{caption}

\usepackage{tabularx}

\usepackage{textcomp}
\usepackage{booktabs}

\usepackage{textgreek}
 
\definecolor{codegreen}{rgb}{0,0.6,0}
\definecolor{codegray}{rgb}{0.5,0.5,0.5}
\definecolor{codepurple}{rgb}{0.58,0,0.82}
\definecolor{backcolour}{rgb}{0.95,0.95,0.92}

\lstdefinestyle{mystyle}{
  language=C,
  float=tp,
  floatplacement=tbp,
  backgroundcolor=\color{backcolour},   commentstyle=\color{codegreen},
  keywordstyle=\color{magenta},
  numberstyle=\tiny\color{codegray},
  stringstyle=\color{codepurple},
  basicstyle=\ttfamily\footnotesize,
  breakatwhitespace=false,         
  breaklines=true,                 
  captionpos=b,                    
  keepspaces=true,                 
  numbers=left,                    
  numbersep=5pt,                  
  showspaces=false,                
  showstringspaces=false,
  showtabs=false,                  
  tabsize=2,
  xleftmargin=1.5em, 
}

\usepackage{balance}

\usepackage{tikz}
\newcommand{\tikzxmark}{%
\tikz[scale=0.23] {
    \draw[line width=0.7,line cap=round] (0,0) to [bend left=6] (1,1);
    \draw[line width=0.7,line cap=round] (0.2,0.95) to [bend right=3] (0.8,0.05);
}}
\newcommand{\tikzcmark}{%
\tikz[scale=0.23] {
    \draw[line width=0.7,line cap=round] (0.25,0) to [bend left=10] (1,1);
    \draw[line width=0.8,line cap=round] (0,0.35) to [bend right=1] (0.23,0);
}}

\newcommand{\tikzcmarkbf}{%
\tikz[scale=0.23] {
    \draw[line width=1.3,line cap=round] (0.25,0) to [bend left=10] (1,1);
    \draw[line width=1.3,line cap=round] (0,0.35) to [bend right=1] (0.23,0);
}}

\newcommand{\tikzhalfcmark}{%
\tikz[scale=0.23] {
    \draw[line width=0.7,line cap=round] (0.25,0) to [bend left=10] (1,1);
    \draw[line width=0.8,line cap=round] (0,0.35) to [bend right=1] (0.23,0);
    \draw[line width=0.7,line cap=round] (0.2,0.95) to [bend right=3] (0.8,0.05);
}}
\usepackage{soul}

\newcommand{\sys}{Yala\xspace}

\newcommand{\snic}{SmartNIC\xspace} 
\newcommand{\snics}{SmartNICs\xspace} 
\newcommand{\bluef}{BlueField-2\xspace}
\newcommand{\bftwo}{BF-2\xspace}
\newcommand{\nic}{NIC\xspace}
\newcommand{\nics}{NICs\xspace}
\newcommand{\slomo}{SLOMO\xspace}

\newcommand{\sota}{state-of-the-art\xspace}
\newcommand{\membench}{mem-bench\xspace}
\newcommand{\regexbench}{regex-bench\xspace}
\newcommand{\comprebench}{compression-bench\xspace}

\newcommand{\eg}{{\it e.g.,\ }}

\newcommand{\ie}{{\it i.e.,\ }}

\newcommand{\aka}{{\it a.k.a.\ }}

\graphicspath{{figures/}}

\begin{document}

\title{Performance Prediction of On-NIC Network Functions with Multi-Resource Contention and Traffic Awareness}

\author{Shaofeng Wu}
\authornote{Equal contribution. The work was done when Qiang was with City University of Hong Kong.}
\email{sfwu22@cse.cuhk.edu.hk}
\orcid{0009-0006-2007-990X}
\affiliation{%
  \institution{The Chinese University of Hong Kong}
  \city{Hong Kong SAR}
  \country{China}
}

\author{Qiang Su}
\authornotemark[1]
\email{jacksonsq97@gmail.com}
\orcid{0000-0002-4482-6248}
\affiliation{%
  \institution{The Chinese University of Hong Kong}
  \city{Hong Kong SAR}
  \country{China}
}

\author{Zhixiong Niu}
\email{zhniu@microsoft.com}
\orcid{0000-0001-6947-9740}
\affiliation{%
  \institution{Microsoft Research}
  \city{Beijing}
  \country{China}
}

\author{Hong Xu}
\email{hongxu@cuhk.edu.hk}
\orcid{0000-0002-9359-9571}
\affiliation{%
  \institution{The Chinese University of Hong Kong}
  \city{Hong Kong SAR}
  \country{China}
}

\begin{abstract}

Network function (NF) offloading on \snics has been widely used in modern data centers, offering benefits in host resource saving and programmability. 
Co-running NFs on the same \snics can cause performance interference due to contention of onboard resources. 
To meet performance SLAs while ensuring efficient resource management, operators need mechanisms to predict NF performance under such contention. 
However, existing solutions lack \snic-specific knowledge and exhibit limited traffic awareness, leading to poor accuracy for on-\nic NFs. 

This paper proposes \sys, a novel performance predictive system for on-\nic NFs. 
\sys builds upon the key observation that co-located NFs contend for multiple resources, including onboard accelerators and the memory subsystem. 
It also facilitates traffic awareness according to the behaviors of individual resources to maintain accuracy as the external traffic attributes vary. 
Evaluation using \bluef \snics shows that \sys improves the prediction accuracy by 78.8\% and reduces SLA violations by 92.2\% compared to \sota approaches, and enables new practical usecases.

\end{abstract}


 
\maketitle
\section{Introduction}
\label{sec:introduction}
\snics have been prevalent in modern data centers to deploy diverse network functions (NF) due to their benefits in programmability and host resource saving~\cite{clara21,clicknp16,offpath22,bf2perf21,moore97,lognic23,e319,uno17,pipedevice22}. They typically integrate heterogeneous onboard resources, such as SoC cores and domain-specific hardware accelerators, to cater to various NF demands. Moreover, vendors are developing increasingly resourceful \nics to meet evolving offloading needs~\cite{bf31,ipu,pensando}. To fully leverage these resources, it is common practice to co-locate multiple NFs on the same \nic~\cite{e319,meili23,fairnic20,uno17}.

Unfortunately, sharing \snic resources among co-located NFs may lead to contention and performance degradation, posing challenges in maintaining the SLAs. One potential solution is to implement stringent resource isolation mechanisms on \snics. Yet, previous work on this front requires specific \nic hardware architecture support and substantial rewriting of NF programs to accommodate new isolation abstractions~\cite{fairnic20,osmosis24}, thus limiting their practical deployment. 
Consequently, developers still need extensive hand-tuning to ensure simultaneous SLA fulfillment for co-located NFs, which is time-consuming and error-prone~\cite{e319,fairnic20,picnic20}.

Ideally, if operators can predict the performance drop an NF will suffer before actually co-running it with other NFs on the same \nic, they can make better resource management decisions on existing infrastructure and NF implementations. 
Concretely, \snic platforms~\cite{e319,uno17,exoplane23,meili23} and SmartNIC-assisted clouds~\cite{awslam,db23,googlecloud} can maximize NF co-locations in \snic offloading while minimizing SLA violations, which correspond to lower total cost of ownership (TCO) for providers and better experience for tenants, and enjoy faster diagnosis and reasoning of on-NIC NF performance compared to slow manual analysis~\cite{clara21,lognic23}.
To achieve these, we need a systematic \textit{on-\nic} NF performance prediction framework, which entails two new challenges. 

\setlength{\tabcolsep}{3pt}
\begin{table}[t]
    \centering
    \small
    \begin{tabular*}{1\linewidth}{@{\extracolsep{\fill}}llll}
        \toprule
    Network Function & Accelerator & \textbf{T} & Framework \\ 
    \midrule
    FlowStats~\cite{slomo20,clara21}& None & \tikzcmark & Click\\
    IPRouter~\cite{slomo20,clara21}&  None & \tikzxmark & Click\\
    IPTunnel~\cite{click2000}&  None & \tikzcmark & Click\\
    NAT~\cite{e319,exoplane23,clara21}& None & \tikzcmark & Click\\
    FlowMonitor~\cite{e319,metron18}& Regex & \tikzcmark & Click\\
    NIDS~\cite{e319,slomo20}&  Regex & \tikzcmark & Click\\
    IPComp Gateway~\cite{comprbf,nfp17}& Regex, compression & \tikzcmark & Click\\  
    ACL~\cite{exoplane23,dpdk_pl} & None & \tikzxmark & DPDK\\ 
    FlowClassifier~\cite{dpdk_pl,opennetvm16} & None & \tikzcmark & DPDK\\ 
    FlowTracker~\cite{doca1} & None & \tikzcmark & DOCA\\ 
    PacketFilter~\cite{doca1} & Regex & \tikzcmark & DOCA\\ 
        \bottomrule
    \end{tabular*}
    \captionof{table}{Typical NFs and accelerators they require from \snics. Common resources (CPU, memory, and NIC subsystems) are not shown. \textbf{T} means that the NF performance heavily depends on the traffic attributes. The regex-based NFs use the same rule set from~\cite{l7filter}. The last column indicates the programming framework we use to implement each NF.}
    \vspace{-5mm}
    \label{tab:nfs}
\end{table}
\setlength{\tabcolsep}{6pt}

First, on-\nic NFs often utilize diverse onboard resources including memory and various hardware accelerators, making it common that contentions occur across heterogeneous resources. 
Prior work on NF performance prediction has primarily focused on memory subsystem contention as the sole source of performance interference for on-server NFs~\cite{slomo20,bubbleup11,dobre12}.
We showcase that the state-of-the-art \slomo~\cite{slomo20} encounters high prediction errors when co-locating NFs contend for both memory and regex accelerator on a \bluef \snic, with $\sim$20\% in the median and $\sim$60\% in the worst case (\cref{moti:multi-resource}). 
The community lacks a clear understanding of (1) the impact of contention on individual domain-specific accelerators and (2) the overall effect of multi-resource contention on performance. 

Second, NF performance is heavily influenced by traffic attributes such as flow counts and payload features, which change dynamically for each NF. 
Current frameworks often either assume fixed traffic attributes~\cite{lognic23,clara21}, or can only deal with a limited range of variations in these attributes (\eg 20\% in flow counts in~\cite{slomo20}).  

In this paper, we propose \textbf{\sys}, a new performance prediction framework that explicitly considers multi-resource contention and dynamic traffic attributes. 
\sys conducts offline profiling of on-\nic NFs to collect their performance under diverse synthetic contention levels and traffic attributes. Leveraging these profiles, \sys trains a contention- and traffic-aware model for each NF, which is then used to predict the NF's performance before its deployment, facilitating placement and other management decisions. 
We build \sys for SoC \snics due to their ease of programmability (\eg DPDK and Click support), 
and tackle the above technical challenges by leveraging critical characteristics of on-\nic NFs. 

\noindent{\bf Multi-resource contention modeling.}
\sys's key idea here is to independently model individual resource contention and integrate these per-resource models together. We identify hardware accelerators and memory subsystems as primary sources of contention for on-NIC NFs. 
For accelerators, we find that it is a common design for NFs to interact with them through their own queues which are coordinated by round-robin scheduling. 
This inspires us to take a white-box approach and propose a queueing-based contention model.  
Memory subsystem contention can be modeled using a black-box ensemble-based ML model following existing work~\cite{slomo20}. 
Then, to capture the end-to-end effect of each resource, we introduce \textit{execution-pattern-based composition}. 
This makes intuitive sense because how each resource and its contention affects the overall performance critically depends on whether NF runs as a pipeline or in a run-to-completion fashion.

\noindent{\textbf{Traffic-aware modeling.}} On top of multi-resource contention, \sys employs \textit{traffic-aware augmentation} to integrate knowledge of traffic attributes into per-resource models.
Generally speaking, this can be done by feeding \textit{traffic attributes}, \eg flow count and packet size, as additional features to per-resource models.
Specifically, for accelerators, we can leverage the white-box nature of the queueing-based model and represent key model parameters as a function of traffic attributes; for memory subsystem which has a blackbox model, we simply fuse traffic attributes with performance counters as features to extend the model. 
In addition, to curb the high profiling cost caused by the introduction of traffic attributes especially for black-box memory models, \sys adopts \textit{adaptive profiling} to prune attribute dimensions and enforce targeted sampling at performance-critical ranges of the attributes.

We implement \sys in C and Python, leveraging typical offline profiling tools~\cite{perf,stressng,mbw,RXPbench} and~\texttt{sklearn}~\cite{sklearn}, and evaluate it on 9 common NFs using \bluef \snic. Our code is open source anonymously at~\cite{yalagithub}. Our testbed evaluation shows that \sys achieves accurate NF throughput predictions under multi-resource contention and varying traffic attributes, with an average error of 3.7\% across NFs which corresponds to 78.8\% improvements compared to \sota \slomo. 
As new usecases, we also illustrate that in NF placement, \sys can reduce SLA violations by 88.5\% and 92.2\% compared to greedy approaches~\cite{e319,meili23} and \slomo, and in performance diagnosis it can deliver higher accuracy in identifying bottlenecks for on-\nic NFs.
\section{Background and Motivation}
\label{sec:motivation}
We start by presenting the brief background of network function resource contention on \snics, followed by the unique challenges of developing a contention-aware performance prediction framework.

\subsection{Background}
\label{subsec:nf_intro}
\snics have been widely used to offload various network functions (NFs) in modern data centers, mainly for their benefits in host resource saving and energy efficiency~\cite{clara21,e319,clicknp16,azure18,meili23,uno17}. 
The NFs leverage the onboard domain-specific hardware accelerators to achieve high throughput and low latency~\cite{e319,exoplane23,slomo20,clara21}. 
We showcase some typical NFs seen across prior work~\cite{e319,exoplane23,slomo20,clara21} and the types of resources they need in Table~\ref{tab:nfs}. 
Here Flow Monitor, NIDS, and PacketFilter require the regex accelerator for packet inspection and payload scanning, {and IPComp Gateway requires both the regex and compression accelerators}. 

\begin{figure}[t]
    \centering
    \small
    \includegraphics[width=1\linewidth]{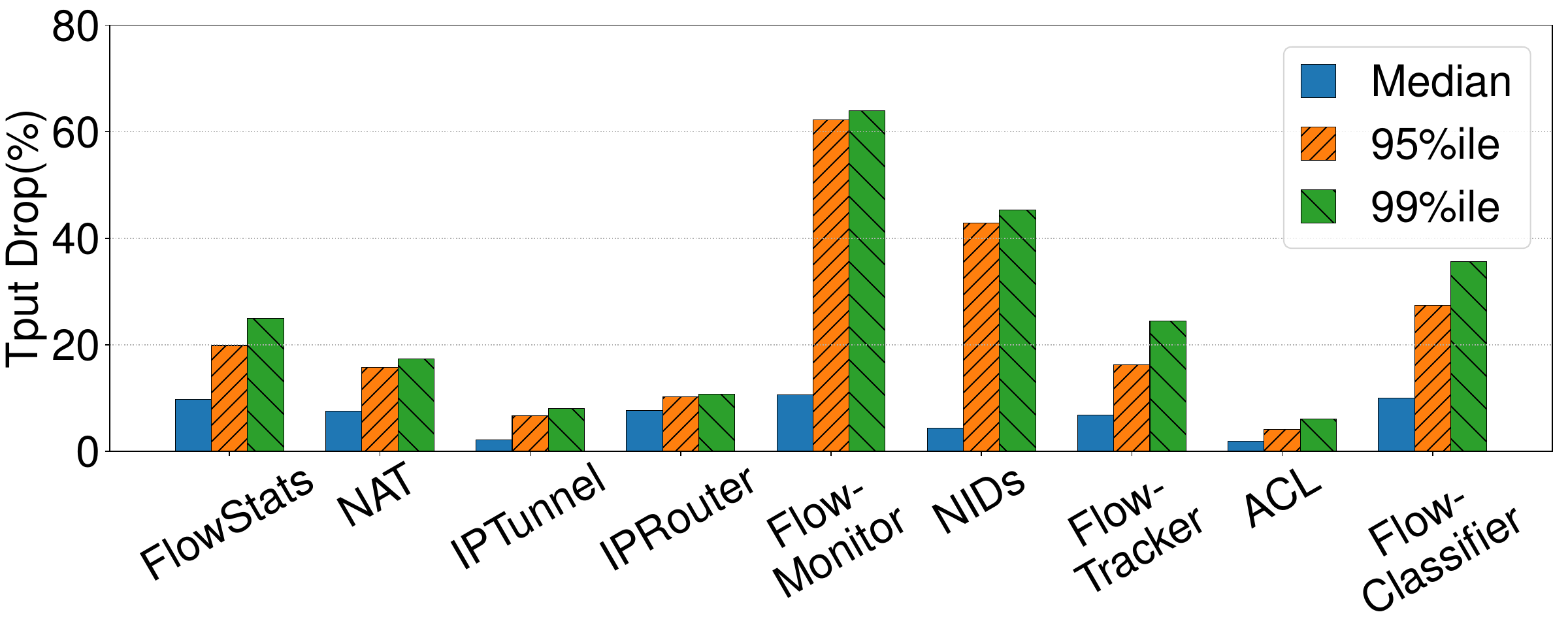}
    \caption{Throughput drop ratios of some NFs from Table~\ref{tab:nfs} under resource contention when co-located with at most 3 other random NFs.}
    \Description[]{}
    \label{fig:coten}
\end{figure}

\noindent{\textbf{Contention degrades performance}}
Recently, co-running multiple NFs on the same \snic has become more common to improve utilization~\cite{e319,meili23,fairnic20,uno17}. 
This can lead to performance degradation due to contention for shared resources. 
To demonstrate this effect, we profile the throughput drop of 9 typical NFs from Table~\ref{tab:nfs} when they co-locate with other NFs.
For each target NF, up to three other NFs are randomly selected (from Table~\ref{tab:nfs})\footnote{Some NFs require minimum two cores, while one \bluef has eight cores in total.}.  
Each NF is given two {dedicated} cores while sharing the memory subsystem and hardware accelerators due to the lack of hardware- or system-level isolation support on current \snics. 
Traffic profiles for all NFs consisted of 16K flows of 1500B packets, with flow sizes following the uniform distribution (further details in \cref{subsec:methodology}). 
For NFs processing payloads with regular expressions, we use \texttt{exrex}~\cite{exrex} to generate packet payloads. 
Note the packet arrival rates are set sufficiently high for all NFs to ensure it is not causing throughput drop. 
We measure the throughput drop ratio against the baseline when the target NF runs alone with two CPU cores, the entire memory and hardware accelerators. 
Figure~\ref{fig:coten} depicts the statistics of throughput drop ratios. 
We can see that when co-running with different (numbers and combinations of) NFs, resource contention can cause {4.2\% to 62.2\%} throughput drop at the 95\%ile, and {1.9\% to 10.6\%} at the median.

\subsection{Challenges}
Modeling and predicting NF performance under resource contention is therefore of paramount importance for many management tasks~\cite{slomo20,lognic23,bubbleup11,dobre12}, and some prior work~\cite{slomo20,bubbleup11,dobre12} has investigated this problem in network function virtualization where NFs run on commodity servers.
An immediate question is, what makes contention-aware performance prediction different in the context of \snics? 
We now highlight two unique challenges which are not well addressed in past efforts.
Note all experiments in this section use the \bluef (\bftwo) \snic. 

\subsubsection{Multi-Resource Contention}
\label{moti:multi-resource}
We've seen that NFs on \snic utilizes multiple heterogeneous onboard resources.
Prior work, however, has only considered contention of the memory subsystems~\cite{slomo20,bubbleup11,dobre12}, missing the contention on other hardware accelerators. Their effectiveness as a result is tainted in the context of \snics. 

To empirically substantiate our argument, {we co-run FlowMonitor with up to three competing NFs chosen randomly from Table~\ref{tab:nfs} on one \bftwo. 
The traffic profiles are identical to the one in Figure~\ref{fig:coten}.} 
We use SLOMO~\cite{slomo20} as the \sota memory-based prediction model and develop a new model for the regex accelerator due to lack of existing models (details in \cref{design:reg_model}).

We first train our single-resource models for FlowMonitor which uses regex accelerator in addition to CPU and memory, and validate their effectiveness under single-resource contention. 
We build two synthetic NFs, \membench and \regexbench\footnote{We build synthetic NFs for three main purposes: 1) collecting training data, 2) exploring insights that support our design choices, and 3) microbenchmarks. For example, regex-bench is purpose-built to have negligible memory subsystem usage but extensive regex accelerator usage, and we rely on it to investigate the contention behavior in regex accelerator (\mbox{\cref{design:reg_model}}). For evaluations on end-to-end accuracy (\mbox{\cref{eva:acc}, \cref{eva:acc_multi}, \cref{eva:acc_traffic}}) and use cases (\mbox{\cref{eva:usecase}}), we employ real NFs from Table~\mbox{\ref{tab:nfs}} instead.}, to assert controllable memory and regex contention, respectively, for generating training data (details in \cref{sec:impl}). 
Following SLOMO, we also collect data from our \bftwo's performance counters at runtime (e.g., memory read/write rates) as the model input.
Absolute percentage error against FlowMonitor's true throughput under single-resource contention is used as the comparing metric. 
Our models achieve the same $<$10\% average prediction error for memory- and regex-only contention as reported in the \slomo paper~\cite{slomo20}. 

\begin{figure}[t]
    \centering
    \small
    \includegraphics[width=1\linewidth]{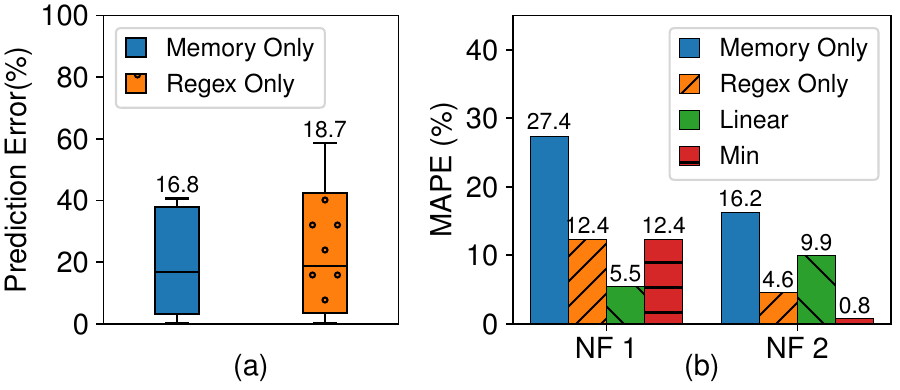}
    \caption{Prediction errors (absolute percentage error) of FlowMonitor's throughput using single-resource models. (a) Box and whisker plot of using only the memory-based \slomo model or a regex-based model (\cref{design:reg_model}). We show the median error on top of each error box. (b) Mean average percentage error (MAPE) of sum and min composition of single-resource models. NF1 and NF2 adopt run-to-completion and pipeline resource usage pattern respectively. 
    }
    \Description[]{}
    \label{fig:multi_resource}
\end{figure}

Then we apply these models directly to the multi-resource contention scenario as said before, where co-locating NFs as a whole contend for both memory and regex accelerator and nothing else. 
Figure~\ref{fig:multi_resource}(a) shows that prediction error now increases to $\sim$20\% in the median and reaches $\sim$60\% in the worst case, indicating that only considering one resource is wildly inaccurate.

In addition, NFs exhibit diverse execution patterns when utilizing these resources.
For example, one NF may run in a \textit{pipeline} manner for high throughput, while another may wait for the completion of dispatched requests to ensure low average latency (\textit{run-to-completion})~\cite{lognic23,RXPbench}. 
This makes \textit{composition} of single-resource models, a strawman solution for multi-resource prediction, inaccurate.

To explore this, we analyze two simple composition approaches:
(1) sum composition, which adds up the predicted throughput loss from each model~\cite{nnmeter,lognic23}, and (2) min composition, which uses the maximum predicted throughput loss as the final output~\cite{e319,flextoe22}. 
Figure~\ref{fig:multi_resource}(b) presents the results of these two approaches in the same setting as Figure~\ref{fig:multi_resource}(a).
We observe that while composition models reduce error, they do not guarantee optimal accuracy across all NFs. 
For NF1 with run-to-completion, sum composition works better, but its error is significant ($\sim$17\%) for the pipeline NF2. 
The key reason is that the resource contention impact on end-to-end throughput varies by NF execution patterns. In pipeline-based NFs, throughput is constrained by the slowest stage on which resource contention causes the most significant performance interference compared. 
In contrast, for run-to-completion NFs, contention on different resources uniformly impacts the end-to-end throughput. 

To quickly recap, NFs on \snics can experience contention across multiple resources, and its impact on performance differs according to the execution patterns. 
Current systems consider only single-resource contention, which results in substantial prediction inaccuracies. 

\subsubsection{Traffic Attributes}
\label{moti:traffic}
\begin{figure}[t]
    \centering
    \small
    \includegraphics[width=1\linewidth]{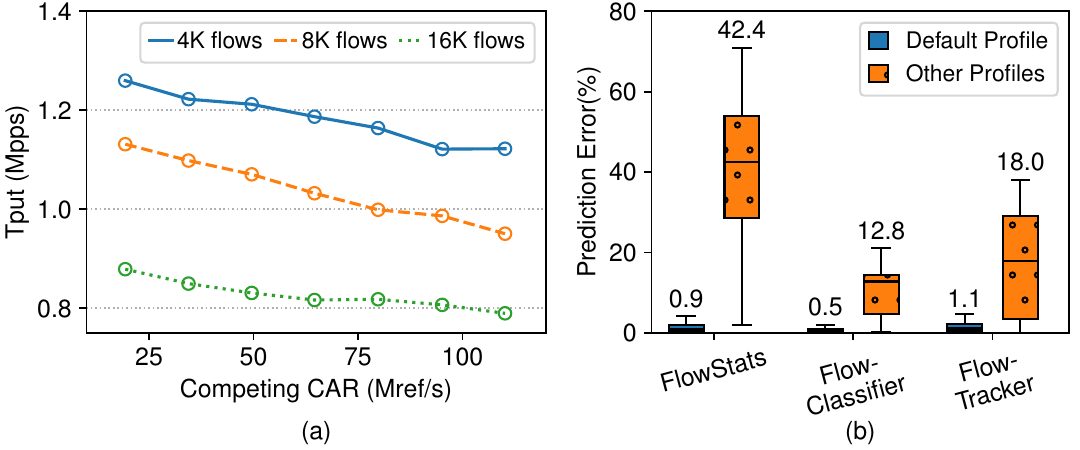}
    \caption{(a) FlowStats's throughput when the competitor's cache access rate (CAR) changes in three distinct traffic profiles. CAR is the sum of the cache read and write rates obtained from the hardware performance counters on \bluef. (b) Distribution of prediction errors after adapting the model to different traffic profiles. We show the median error on top of each error box.}
    \Description[]{}
    \label{fig:flow_adapt}
\end{figure}

An NF's performance also depends on certain traffic attributes, such as number of flows, payload characteristics, etc., in many cases~\cite{lognic23,slomo20,clara21}. 
To see this, we measure FlowStat's throughput when co-located with \membench, and vary \membench's cache access rates (CAR). 
Figure~\ref{fig:flow_adapt}(a) shows that FlowStat's throughput drops differently in different traffic profiles as \membench's CAR increases, implying that a traffic-agnostic model inevitably leads to high prediction errors when adapting to new traffic profiles.

Figure~\ref{fig:flow_adapt}(b) empirically confirms the intuition above for existing work. 
Here we look at three target NFs: FlowStats, FlowClassifier, and FlowTracker. 
Each of them is co-located with \membench on a single \bftwo. 
We use the same default traffic profiles of 16K flows to train three models for each target NF following \slomo just as in the experiments before.
We then test them under changing traffic attributes by generating 100 distinct traffic profiles with random number of flows up to 500K.
It is clear from Figure~\ref{fig:flow_adapt}(b) that prediction error increases dramatically when the traffic behavior deviates from the default profiles that the models have seen. 
Note \slomo does consider the number of flows in its prediction, but can only handle a small degree of deviation from the training data as we shall detail in \cref{subsec:methodology} and \cref{eva:acc_traffic}.

\section{\sys Overview}
The challenges in \cref{sec:motivation} pose two fundamental questions on accurate performance prediction for on-\nic NFs, which drive \sys's design:
\begin{enumerate}
    \item How to model the impact of multi-resource contention on NF performance?
    \item How to integrate traffic attributes to contention-aware performance prediction models? 
\end{enumerate}

We develop {\sys}, a framework for accurately predicting on-\nic NF performance with multi-resource contention and varying traffic profiles.
To address the first design question, \sys adopts a ``divide-and-compose'' approach: it builds up individual \textit{per-resource contention models} (\cref{design:resource_model}) for both hardware accelerators and memory subsystem to separately model their impact on throughput, and applies \textit{execution-pattern-based composition} (\cref{design:resource_compose}) to faithfully capture the end-to-end effect of contention.
Then in response to the second design question, \sys introduces \textit{traffic-aware augmentation} (\cref{design:augmentation}) techniques to integrate various traffic attributes into per-resource models, and develops an \textit{adaptive profiling} (\cref{design:adaptive}) method to balance the soaring profiling costs (due to the extra dimensions of traffic attributes) with model quality.
Taken together, during online prediction, \sys takes the contention level of competing NFs and traffic attributes of the target NF as input to the per-resource models and compose the results based on NF's execution pattern to obtain the final prediction. 
Consistent with prior work~\cite{slomo20,bubbleup11,lognic23}, \sys does not require knowledge of or access to NF source code.

\section{Multi-Resource Contention Modeling}
\label{sec:models}
We now present the design insights and details of \sys's multi-resource contention modeling. 
Note we are interested in the NF's maximum throughput assuming the arrival rate is high enough, which represents the NF's capability and is consistent with prior work~\cite{slomo20,bubbleup11,clara21}.

\subsection{Per-Resource Models}
\label{design:resource_model}
An on-\nic NF consumes onboard CPU, memory subsystem (cache and main memory), hardware accelerators, and NIC~\cite{clara21,fairnic20,picnic20,lognic23}. For CPU, given common deployment practice~\cite{resq18,nfp17,opennetvm16}, we perform core-level isolation for co-located NFs so CPU contention does not happen. 
Although some prior work has discussed potential isolation issues for \nics on a server~\cite{picnic20}, we do not encounter this problem as on-\nic NFs leverage powerful hardware-based flow table on \snics~\cite{ovsdpdk}. 
Thus, we focus on contention on hardware accelerators and memory subsystem here, assuming fixed traffic attributes.
Notice the per-resource modeling effectively derives the NF's throughput on one given resource only without accounting for other resources, and may not equal to the overall throughput.

\subsubsection{Hardware Accelerators}
\label{design:reg_model}
\label{overview:cont_regex}
At first glance, modeling hardware accelerator contention seems not much different from existing design for memory contention~\cite{slomo20,bubbleup11}.
That is, one can use an accelerator's performance counters to quantify NF's contention level as the input, and employ an ML model to predict throughput.
This is infeasible, however, because current \snic accelerators do not expose fine-grained performance counters~\cite{bf21,regexbf,bf2perf21,perf,bf2forum}.
We thus propose a general \textit{queue-based} white-box approach for hardware accelerators.

\noindent{\textbf{Contention behavior in hardware accelerators.}} 
We start by analyzing the accelerator's contention behavior which our modeling is based upon. 
Without loss of generality, we use the widely-used regex accelerator~\cite{RXPbench,doca1,bf2forum} as the target of discussion hereafter.

In practice, NFs utilize onboard accelerators via the corresponding queue systems~\cite{DPDK,doca1}.
For example, an NF establishes request queues and enqueues/dequeues operations to/from a regex accelerator~\cite{DPDK,RXPbench,doca1}. This queue-based interface unifies the interaction with specialized accelerators and applies to many \snics~\cite{pensando,bf21,bf31} and beyond~\cite{KKA16,rdma-qp}. 
Understanding the queue system behavior is then crucial for modeling accelerator contention.

\noindent{\it Setup.} We write a synthetic Click NF called regex-NF that utilizes regex accelerator to scan packet payloads.
regex-NF's packet arrival rate is high enough to ensure maximum throughput, and it is tested with different match-to-byte ratios (MTBR).\footnote{Match-to-byte-ratio (MTBR) refers to how many matches against a regex ruleset is contained in each byte of the payload. A higher MTBR reflects more regex matches with in each unit of packet payload and longer processing time for a packet (\cref{subsec:methodology}).}
To vary contention level, we adjust the co-running \regexbench's arrival rate. 

\begin{figure}[t]
    \centering
    \small
    \includegraphics[width=1\linewidth]{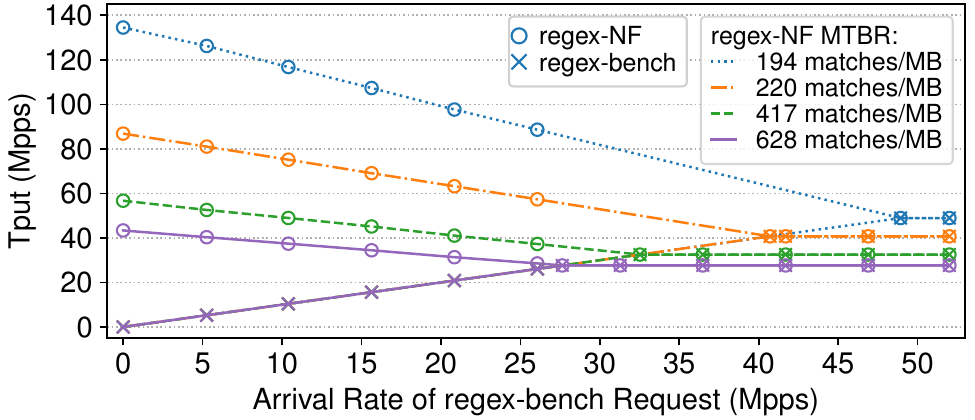}
    \caption{Throughput of co-running synthetic pattern-matching ``regex-NF'' and \regexbench as a function of arrival rate of \regexbench. In each setting, regex-NF and \regexbench reach an equilibrium throughput, \eg, with MTBR of 194 maches/MB for regex-NF, they both obtain 48.9 Mpps at equilibrium.}
    \Description[]{}
    \label{fig:cont_src_reg}
\end{figure}

\noindent{\it Observation.} We depict the throughput results in Figure~\ref{fig:cont_src_reg} and make two interesting observations. 
\textit{O1:} First, regex-NF shows \textit{linear} throughput drop as the contention from regex-bench rises. 
\textit{O2:} Second, regex-NF finally reaches the equilibrium throughput without further dropping. 
The equilibrium point clearly varies with \regexbench's MTBR. 

These two observations are very familiar to us as they point to the canonical round-robin (RR) queuing discipline widely used in practice. Indeed, we confirm from~\cite{mlxregex} that our regex accelerator driver's implementation adopts RR for queue-level fairness.
With one queue per each NF which is the setup in our experiment, as \regexbench's request arrival rate 
increases, regex-NF's requests have proportionally less access to the accelerator, resulting in the linear throughput decline.
When contention is high enough that \regexbench's queue is always non-empty when the RR scheduler turns to it, throughput of regex-NF stops dropping 
since its average sojourn time (sum of queuing and processing times) stops increasing~\cite{queue1}.
As they have the same numbers of queues, their equilibrium throughput is the same as seen in Figure~\ref{fig:cont_src_reg}.

\noindent{\textbf{Our approach.}} 
Motivated by the above analysis, the regex accelerator contention can be modeled by RR over multiple request queues with one service node (\ie the accelerator).
Suppose we have $N$ NFs sharing an accelerator, and each $NF_j$ has $n_j$ request queues. 
At equilibrium, the average sojourn time $t$ of requests from each queue is~\cite{queue1}: $t=\sum_{j=1}^{N}n_jt_j$, where $t_j$ represents $NF_j$'s average request processing time. 
For a target $NF_i$, its throughput $T_i$ can be represented as the sum of throughput of all its queues, \ie      
\begin{equation}
\begin{aligned}
T_i &= \frac{n_i}{t} = \frac{n_i}{\displaystyle\sum_{j=1}^{N}n_jt_j} = \frac{n_i}{\displaystyle\sum_{j=1}^{N}\frac{n_j^2}{T_{j,solo}}} , \\
\end{aligned}
\label{eq:regex_cont}
\end{equation}
where $T_{j,solo}$ represents its regex processing throughput (in pps) when $NF_j$ runs solo. 
Clearly when $n_i=n$ for all NFs, they have the same (equilibrium) throughput $T_i$.

Now to use Equation~\eqref{eq:regex_cont} for a new NF, we need to infer $n_j$ and $T_{j,solo}$ without any knowledge of the NF. 
Recall $T_{j,solo}$ is throughput on the regex accelerator only, which may or may not equal to end-to-end throughput if the NF is bottlenecked on other resources or follows run-to-completion.
So to estimate them accurately, we again co-run the NF with \regexbench and set \regexbench's request processing time and match rate to be high enough to ensure that at equilibrium, the NF spends most of its time on regex.
We then collect two sets of equilibrium throughput data to solve for $n_j$ and $T_{j,solo}$ since \regexbench's parameters are known.

We verify Equation~\eqref{eq:regex_cont} with empirical results of various regex-based NFs, which show that our approach is accurate with 1.3\% error on average.

\noindent{\bf Other accelerators.}
Our approach here directly applies to other hardware accelerators, \eg compression and crypto accelerator, which also uses round-robin based queues\cite{comprbf,pkabf}. 

\subsubsection{Memory Subsystem}
\label{design:mem_model}
\label{design:contetnion_mem}
Memory subsystem contention has been studied in existing work~\cite{slomo20,bubbleup11,dobre12} which finds that the contention-induced throughput drop can be modeled as a \textit{piece-wise linear} function of performance counters. 
Thus we follow \slomo's gradient boosting regression (GBR) method which is \sota, using 7 performance counters as input features.
Note that we overcome the fixed-traffic limitation of GBR by integrating traffic attributes to it in \cref{sec:tfaware}.

\subsection{Execution-Pattern-Based Composition}
\label{design:resource_compose}

We now discuss how to composite the per-resource models for deriving end-to-end throughput. 

\noindent{\textbf{Observations.}}
We analyze two typical execution patterns of NFs: \textit{pipeline} and \textit{run-to-completion}~\cite{lognic23,RXPbench}.
In the following discussion, we define a \textit{stage} as a processing block that will only utilize one resource type.
Considering a packet received by a pipeline NF, or p-NF, and a run-to-completion NF, or r-NF: 
for p-NF, the packet waits at the first stage until its predecessor \textit{enters} the second stage;  
for r-NF, the packet waits until the predecessor \textit{leaves} the last stage.

\begin{figure}[t]
    \centering
    \small
    \includegraphics[width=1\linewidth]{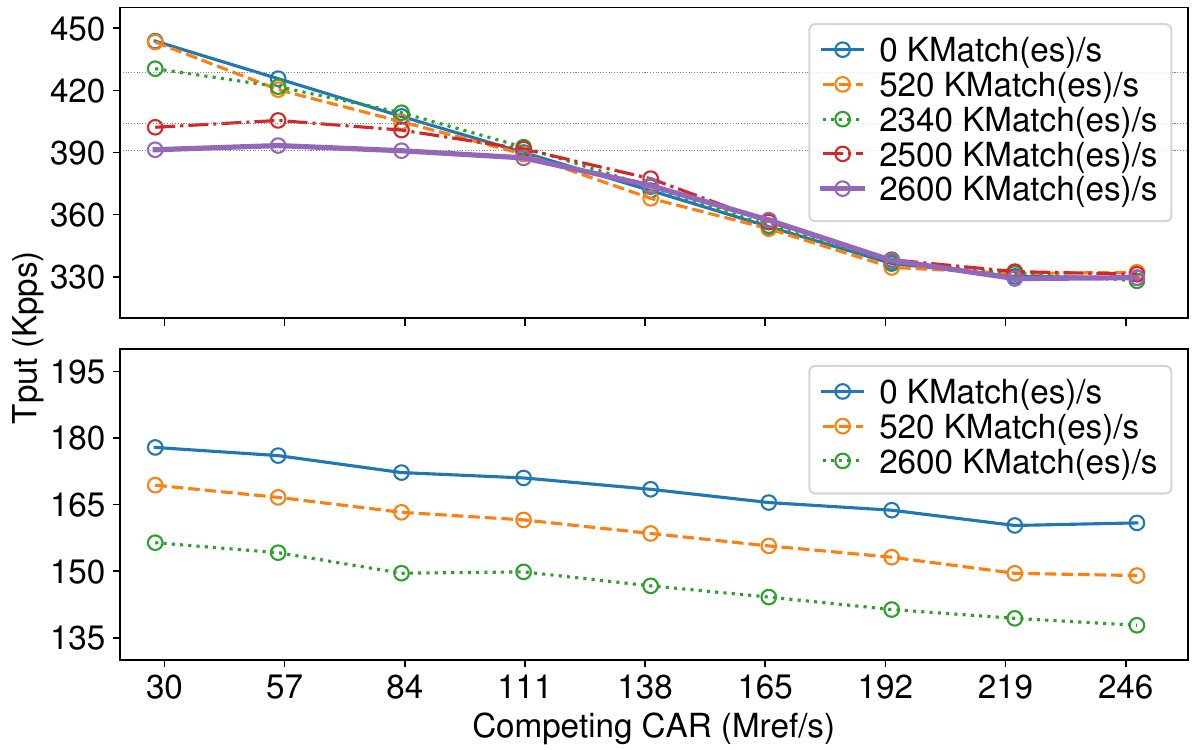}
    \caption{Throughput of two synthetic Click NFs that use pipeline (top) and run-to-completion (bottom) as a function of competing CAR in memory subsystem and match rate in regex accelerator. }
    \Description[]{}
    \vspace{-1mm}
    \label{fig:pattern_tput}
\end{figure}

Figure~\ref{fig:pattern_tput} presents the throughput of a synthetic p-NF (top) and r-NF (bottom) under different levels of memory and regex accelerator contention.
We observe that: \textit{O1.} the p-NF's throughput stays unchanged when memory contention is low and regex contention is high.
For example, the throughput stays at $\sim$400~Kpps when competing cache access rate (CAR) 
is less than $\sim$100 Mref/s and competing match rate (product of throughput and MTBR) is 2500 Kmatches/s. 
This is because the throughput of a pipeline equals that of its slowest stage --- regex matching in this case, making it insensitive to memory subsystem contention.
\textit{O2:} Second, for the r-NF, we observe that throughput drop is a \textit{monotonically decreasing} function of both competing CAR and regex match rate, indicating that throughput drop is always caused by the compounded contention.  

\noindent{\textbf{Our approach.}} 
The main goal here is to derive a composing function that takes in execution pattern and per-resource throughput drop $\Delta T_k, 1\leq k\leq r$ (given by per-resource models) as input, and produces the end-to-end throughput drop caused by contention in $r$ resources.

\textit{Pipeline:} Based on \textit{O1}, end-to-end throughput (denoted as $T$) of a p-NF can be calculated as:
\begin{equation}
\begin{aligned}
T &= T_{solo} - \max(\Delta T_1, ..., \Delta T_r) ,\\
\end{aligned}
\label{eq:pipeline}
\end{equation}
where $T_{solo}$ is the NF's throughput when running solo. 

\textit{Run-to-completion}: Based on \textit{O2}, 
we denote the processing time of a packet in each resource without contention as $t_k$ (also the sojourn time), where $1\leq k \leq r$. 
Due to multi-resource contention, the sojourn time of a packet in each resource grows by $\Delta t_k$ as a result of throughput drop $\Delta T_k$. Therefore, the throughput of the r-NF can be represented as:

\begin{equation}
\begin{aligned}
T &= \frac{1}{\displaystyle \sum_{k=1}^{r}(t_k + \Delta t_k)} = \frac{1}{\displaystyle \sum_{j=1}^{r}\left(t_j+\Delta t_j+\sum_{k=1,k\neq j}^{r}t_k\right) - \sum_{j=1}^{r}\sum_{k=1,k\neq j}^{r}t_k} \\
&= \frac{1}{\displaystyle \sum_{j=1}^{r}\frac{1}{T_{solo} - \Delta T_j} - \frac{r-1}{T_{solo}}}.
\end{aligned}
\label{eq:rtc}
\end{equation}

\noindent\textbf{Detecting execution pattern.} 
Without source code access, we resort to a simple testing procedure to detect an NF's execution pattern.
We co-run the NF with our benchmark NFs, and see if Equation~\ref{eq:pipeline} or \ref{eq:rtc} fits its throughput drop better. 
One may also observe the NF's throughput curve similar to Figure~\ref{fig:pattern_tput} to empirically determine if it is a p- or r-NF.

\section{Traffic-Aware Prediction}
\label{sec:tfaware}

Our discussion so far has been limited to fixed NF traffic profiles. 
Now we discuss how to integrate traffic attributes into our models.

\subsection{Traffic-Aware Augmentation}
\label{design:augmentation}
It is obvious that we need to augment the per-resource model with knowledge of traffic attributes, while execution-pattern-based composition is not affected.
To do this, we select three common traffic attributes that impact NF performance based on our experiment results and previous studies~\cite{slomo20,lognic23}: number of flows or flow count, packet size, and match-to-byte-ratio (MTBR) of a packet. 
We denote a traffic profile of 16K flows, 1500B packets and 600 matches/MB payload using a vector $(16000, 1500, 600)$.

\subsubsection{Hardware Accelerators}
\label{design:traffic_tip}
We again start with hardware accelerators, specifically the regex accelerator, as a concrete example.

\noindent{\textbf{Our approach.}} 
A regex-utilizing NF is naturally sensitive to the MTBR of packet payload~\cite{RXPbench}.\footnote{Number of queues is fixed during NF's life cycle per its configuration.} 
It directly impacts the average processing time of the regex requests.

Following notations from Equation~\ref{eq:regex_cont}, the average request processing time of $NF_j$ for a given ruleset can be expressed as: $t_j = \frac{1}{T_{j,solo}} = t_{j,0} + a_j m_j$, where $t_{j,0}$ and $a_j$ are constants, and $m_j$ is the MTBR.  
This equation builds on the intuition that that request processing time grows linearly with number of matches in a packet, as each match induces a constant amount of extra processing time on average on top of the base processing time. 
Plugging into Equation~\ref{eq:regex_cont}, the traffic-aware throughput model of regex accelerator is:
\begin{equation}
\begin{aligned}
T_i &= \frac{n_i}{\sum_{j=1}^{N}n_j^2(t_{j,0} + a_jm_j)}. \\
\end{aligned}
\label{eq:regex_cont_traffic}
\end{equation}

The parameters $t_{j,0}, a_j$ can be easily obtained from linear regression using data from co-running the NF with \regexbench under different MTBRs.

\noindent{\bf Other accelerators.}
Per-resource models for other accelerators can be augmented in a similar two-step approach: (1) analyzing what accelerator-specific traffic attributes affect the NF; (2) representing the average processing time as a function of these accelerator-specific traffic attributes.

\subsubsection{Memory Subsystem}
\label{design:mem_traffic}
\begin{figure}[t]
    \centering
    \small
    \includegraphics[width=1\linewidth]{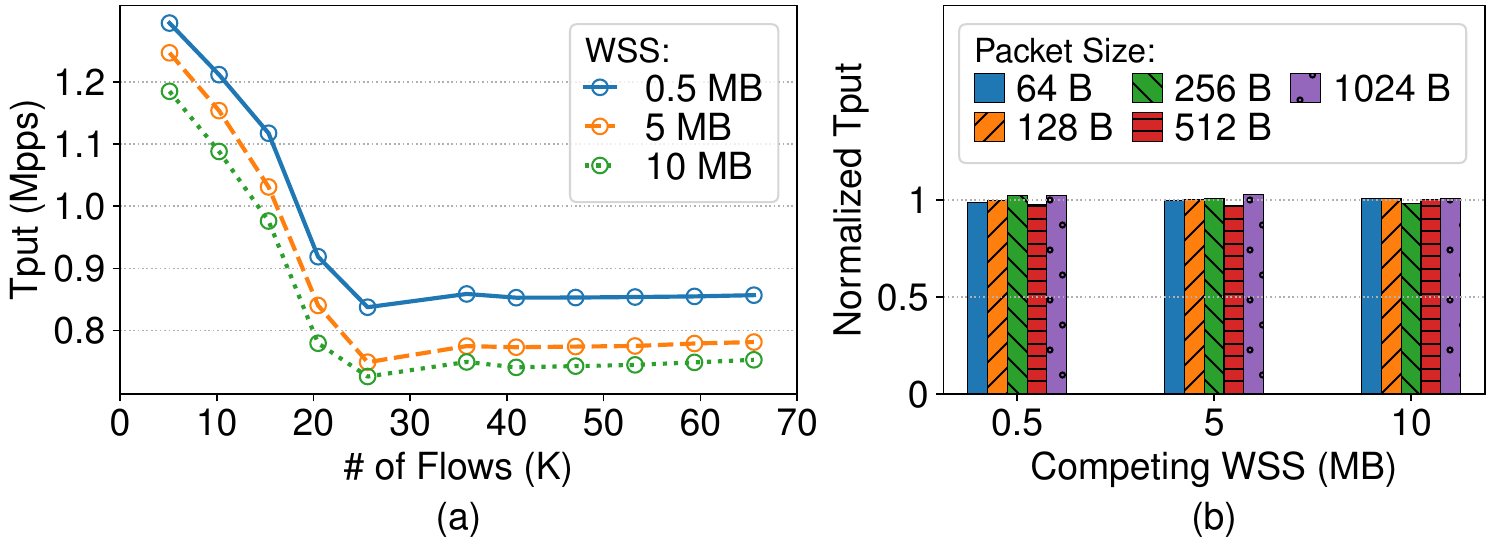}
    \caption{Throughput of FlowStats as a function of traffic attributes. (a) The packet size is 1500 B. (b) The number of flows is 16 K.}
    \Description[]{}
    \label{fig:traffic_insight}
\end{figure}

We co-run FlowStats with \membench with different numbers of flows (other traffic attributes stay the same).
We adjust the working set size of \membench and keep other metrics (\ie CAR and memory access rate) fixed.
Figure~\ref{fig:traffic_insight}(a) reveals that FlowStats experiences significant throughput drop with increasing flow count. 
More interestingly, the throughput curve is also a \textit{piece-wise} function of the number of flows, which is similar to memory contention modeling as mentioned in \cref{design:mem_model} (we shall explain this in \cref{design:adaptive} later). 
Thus, we take a straight-forward approach to add traffic attribute knowledge as extra features to the blackbox GBR model of memory subsystem, by appending the traffic attributes vector to the original input vector of performance counters.

\subsection{Adaptive Profiling}
\label{design:adaptive}
Being traffic-aware brings a new and critical issue: training a traffic-aware model now needs more data due to the higher dimensionality of the input vector compared to a fixed-traffic model. This is especially true for the black-box model for memory subsystem, which already has a high-dimensional input vector.
A naive solution is to repeat the collection process for each unique traffic profile (\aka full profiling), which obviously leads to massive exponentially-growing profiling overheads.
Although random sampling based profiling~\cite{slomo20,bubbleup11} can effectively reduce overhead, it may cause high prediction error due to inadequate coverage of the data space.
The trade-off between profiling cost and model accuracy motivates us to design an optimized profiling method for \sys.

\noindent{\textbf{Observations.}} 
Our design has roots on two key observations.
First, an NF's performance is only dependent on a few traffic attributes. For example, FlowStats is only sensitive to number of flows as in Figure~\ref{fig:traffic_insight}(a), but not packet size as in Figure~\ref{fig:traffic_insight}(b), since it only processes packet header.

Second, for a given traffic attribute, it causes significant performance drops only in a limited range of values; performance shows little changes in other ranges.
Still using FlowStats, when the competing WSS is 10MB, its throughput loses $\sim$30\% for $[1, 20K]$ flows, but remains almost unchanged for $[40,60K]$ flows as in Figure~\ref{fig:traffic_insight}(a). 
This is because FlowStats's hash table grows as a result of growing flow count. Before it fully occupies the last level cache (LLC), increasing the hash table size leads to higher cache miss ratio, which slows down read/write operations and thus throughput. 
After the LLC is saturated, cache miss ratio stays at a fixed level, causing NFs to exhibit constant throughput.
These characteristics also hold in general across NFs as we empirically observe for other NFs in Table~\ref{tab:nfs}. 
This is because generally, (1) NFs typically process either packet headers or payloads, which only depends on several traffic attributes; 
and (2) traffic attributes usually affect performance by changing the size of key data structures in the NF processing logic, \eg the mapping table in NAT~\cite{slomo20,click2000,bubbleup11}, which exhibits the same LLC effect as explained just now.

\noindent{\textbf{Our approach.}} 
Our key idea is to prune irrelevant traffic attribute dimensions, and conduct more sampling for relevant traffic attributes within the critical value ranges where NF performance has salient changes. 
We propose a two-step adaptive profiling algorithm that balances between model accuracy and profiling cost. 

As shown in Algorithm~\ref{alg:adaptive_profile}, we first test whether NF performance is sensitive to a traffic attribute or not. 
Suppose the possible range of an attribute $f$ is $[f_{min},f_{max}]$. We profile the NF's solo throughput with $f$ set to $f_{min}$ and $f_{max}$ respectively while other attributes remain default (lines 8-9). 
Then we compare the throughput difference to determine if $f$ should be added in our model.
Note that the function \textproc{profile\_one()} collects throughput data under a specified configuration (contention level and traffic attribute), and increments the total number of collected samples by one if the configuration has not been profiled.
After pruning the attribute list, we carry out a binary search to adaptively collect profiling data. 
For each call to \textproc{range\_profile()}, we consider the difference between solo throughputs on traffic attribute boundaries. 
If their difference exceeds a certain threshold, we collect $m$ data points at the center of current traffic attribute region (lines 18-22), 
Then the traffic attribute region is split in half, which will serve as the new region for the recursive call of \textproc{range\_profile()}. 

\begin{algorithm}[t]
    \centering
    \small
    \caption{Adaptive profiling.}
    \label{alg:adaptive_profile}
    \begin{algorithmic}[1]
        \State {$nf$: The target NF to be profiled}
        \State {$\mathcal{C}$: The list of performance counters}
        \State {$\mathcal{F}$: The list of traffic attributes}
        \State {$q$, $\epsilon_0$, $\epsilon_1$, $m$ are hyperparameters}
        \Function {adaptive\_profile}{$nf$, $\mathcal{C}$, $\mathcal{F}$}
        \State {\textit{$n$} $\leftarrow$ 0}
        \For {$f$ in $\mathcal{F}$}
            \State {$T_{min}$ $\leftarrow$ \textproc{profile\_one}($nf$, $0$, $f_{min}$, $n$)}  
            \State {$T_{max}$ $\leftarrow$ \textproc{profile\_one}($NF$, $0$, $f_{max}$, $n$)} \Comment{"$0$" represents no contention. $f_{min}$ and $f_{max}$ represent the lowest and highest possible value of $f$ respectively.}
            \If {\textit{$|T_{max}-T_{min}|$} $<$ $\epsilon_0$}
                \State {$\mathcal{F} \leftarrow \mathcal{F} - \{f\}$} \Comment{Prune the traffic attribute.} 
            \EndIf
        \EndFor
        \State {\textproc{range\_profile}($nf$, $\mathcal{F}_{min}$, $\mathcal{F}_{max}$, $n$)} \Comment{$\mathcal{F}_{min}$ and $\mathcal{F}_{max}$ mean all traffic attributes take the lowest and highest possible value respectively.} 
        \State \Return
        \EndFunction
        
        \Function {range\_profile}{$nf$, $\mathcal{F}_{min}$, $\mathcal{F}_{max}$, $n$}
        \State {$T_{min}$ $\leftarrow$ \textproc{profile\_one}($nf$, 0, $\mathcal{F}_{min}$, $n$)}
        \State {$T_{max}$ $\leftarrow$ \textproc{profile\_one}($nf$, 0, $\mathcal{F}_{max}$, $n$)}
        \If {$n \geq q$} \Comment{We reach profiling quota.}
            \State \Return
        \EndIf
        \If {\textit{$|T_{max}-T_{min}|$} $\geq$ $\epsilon_1$} 
            \State{$\mathcal{F}_{mid}$ $\leftarrow$ $\frac{\mathcal{F}_{max}+\mathcal{F}_{min}}{2}$}
            \For {\_ in $m$}
                \State {$\mathcal{C}_{r}$ $\leftarrow$ \textproc{random}()} \Comment{Choose a random contention level to apply on target NF.}
                \State {\textproc{profile\_one}($nf$, $\mathcal{C}_{r}$, $\mathcal{F}_{mid}$, $n$)} 
            \EndFor
            \State {\textproc{range\_profile}($nf$, $\mathcal{F}_{mid}$, $\mathcal{F}_{max}$, $n$)}  
            \State {\textproc{range\_profile}($nf$, $\mathcal{F}_{min}$, $\mathcal{F}_{mid}$, $n$)}
        \EndIf
            \State \Return
        \EndFunction
    \end{algorithmic}
\end{algorithm}

\section{Implementation}
\label{sec:impl}
We implement \sys with $\sim$1600 LoCs in {C and Python}.
\sys employs \texttt{perf-tools}~\cite{perf} and a working set size estimation tool~\cite{wss} to collect performance counters. 
\sys uses \texttt{sklearn}~\cite{sklearn} to construct machine learning models used in per-resource models. 
Our code is open source anonymously at~\cite{yalagithub}. 

\noindent{\textbf{Synthetic benchmarking NFs.}} 
We implement three synthetic NFs called \membench, \regexbench and \comprebench ($\sim$8300 LoCs {in C with DPDK support}) based on open-source benchmark tools~\cite{stressng,mbw,RXPbench,DPDK} to apply configurable levels of contention on memory subsystem, regex, and compression accelerators, respectively.

\noindent{\textbf{Network functions.}} 
We implement common on-\nic NFs with $\sim$3600 LoCs {in C and Click} using frameworks including Click 2.1~\cite{click2000}, DPDK 20.11.6~\cite{dpdk_pl}, and DOCA 1.5-LTS~\cite{doca1}. 
Our \bftwo enables hardware flow table offloading for NFs. 


\section{Evaluation}

We present our evaluation of \sys now. 
The highlights are: 
\begin{enumerate}
    \item[(1)] \textbf{Accuracy:} \sys achieves an average prediction error of 3.7\% end-to-end, with 78.8\% error reduction compared to \sota across NFs (\cref{eva:acc}). Microbenchmarks further show that \sys's design choices on multi-resource contentions and changing traffic attributes are effective (\cref{eva:acc_multi}, \cref{eva:acc_traffic}).
    \item[(2)] \textbf{Usecases:} To see how \sys can be beneficial in practice, we show two concrete usecases where it (1) facilitates resource-efficient placement decisions in NF scheduling, reducing NF SLA violations by 88.5\% and 92.2\% compared to the classical greedy-based approaches and \slomo, and (2) enables fast performance diagnosis for NFs under contention with 100\% accuracy (\cref{eva:usecase}). 
    \item[(3)] \textbf{Overhead:} \sys's offline adaptive profiling reduces profiling cost while maintaining high model accuracy (\cref{eva:micro}). 
\end{enumerate}

\subsection{Methodology}
\label{subsec:methodology}

We employ NVIDIA \bluef (\bftwo) to evaluate \sys. 
A \bftwo \snic has 8 ARMv8 A72 cores at 2.5GHz, 6MB L3 cache, 16GB DDR4 DRAM, dual ConnectX-6 100GbE ports, and hardware accelerators for regex and compression. 
The NF traffic is generated from a client machine with an AMD EPYC-7542 CPU with 32 cores at 2.9GHz and a ConnectX-6 100GbE NIC. 
Both the \bftwo server and client machine are connected to a Mellanox SN2700 switch. 

\noindent{\bf Traffic profiles.}
We employ \texttt{DPDK-Pktgen}~\cite{pktgen} to create various traffic profiles with different attributes, \ie number of flows and packet sizes. 
In addition, we generate packet payloads using \texttt{exrex}~\cite{exrex} with diverse MTBR of conducting regular expression matches for NFs using regex accelerator. 
The rule sets are from~\cite{l7filter}.

\noindent{\bf Baseline.}
\slomo~\cite{slomo20} serves as our baseline. 
For each NF, we train a model using \slomo's gradient boost regression under the default traffic profile, 
which has 16K flows, 1500B packet size, and the MTBR at 600 matches/MB.
If in testing the traffic profile deviates from default, we employ \slomo's sensitivity extrapolation\footnote{``\textit{Sensitivity}'' refers to NF performance as a function of contention level for simplicity~\cite{dobre12,slomo20,bubbleup11,opportunity11}.} to adapt the model (Section~6 in~\cite{slomo20}). 
We validate that our models achieve the same level of prediction error as in~\cite{slomo20}. 

\noindent{\bf Metrics.}
We report mean absolute percentage error (MAPE)~\cite{slomo20,prophet17,bubbleup11} as the main metric of prediction accuracy. 
We also report \textit{$\pm5\%$ Acc.} and \textit{$\pm10\%$ Acc.} to avoid the impact of test set size variations~\cite{nnmeter}.

\subsection{Overall Accuracy}
\label{eva:acc}
We first evaluate the overall prediction accuracy of \sys under multi-resource contention and varying traffic attributes.
Each target NF is co-located with up to three other NFs and we numerate all possible combinations of NFs. 
We also apply 9 distinct traffic profiles for each NF. 
The results are aggregated under all traffic profiles, as shown in Table~\ref{tab:err_real}. 
We can see that \sys averagely exhibits 3.7\% MAPE, 70.8\% \textit{$\pm5\%$ Acc.} and 95.9\% \textit{$\pm10\%$ Acc.}, compared to \slomo's 17.5\% MAPE, 50.0\% \textit{$\pm5\%$ Acc.} and 72.7\% \textit{$\pm10\%$ Acc.}, demonstrating \sys's superior prediction accuracy.
Specifically, \sys has the most significant gains for IPTunnel, FlowMonitor, FlowStats, and NIDS that use multiple resources and/or is sensitive to traffic attributes. 
Meanwhile, \sys achieves the same accuracy as \slomo for ACL because it is very lightweight and insensitive to traffic attributes. 
Note that the high prediction accuracy for such NFs is aligned with that from previous studies~\cite{slomo20,dobre12,bubbleup11}. 

Next, we microbenchmark \sys's major design choices to better understand the benefits they each bring. 

\setlength{\tabcolsep}{3pt}
\begin{table}[t]
    \centering
    \footnotesize
    \begin{tabular*}{1\linewidth}{@{\extracolsep{\fill}}ccccccc}
        \toprule
        \multirow{2}{*}{NF} & \multicolumn{3}{c}{SLOMO} & \multicolumn{3}{c}{\sys} \\ 
        \cmidrule(lr){2-4}\cmidrule(lr){5-7}
          & \makecell{MAPE\\(\%)} & \makecell{$\pm$5\% \\ Acc.(\%)} & \makecell{$\pm$10\% \\ Acc.(\%)} &\makecell{MAPE\\(\%)}& \makecell{$\pm$5\% \\ Acc.(\%)} & \makecell{$\pm$10\% \\ Acc.(\%)} \\ 
          \midrule
        ACL    & 1.3 & 100.0 & 100.0  & 1.2 & 100.0 &  100.0 \\
        NIDS        & 16.2 &  24.3 & 74.3 & 1.5 & 95.9 & 100.0    \\ 
        IPTunnel    & 62.9  & 70.5  & 73.1 &  3.8  & 75.6 & 92.3 \\
        IPRouter & 4.2 &   68.4 &  98.2 & 3.8       & 66.7 & 100.0 \\
        FlowClassifier  &  7.5  & 28.1 &  73.7 &  3.8  &  63.2  &  100.0 \\
        FlowTracker   & 4.9 &  56.1 & 86.0 & 3.9  & 61.4 & 100.0 \\
        FlowStats   & 11.7 & 33.3 & 57.9 & 4.3  & 70.2  & 96.5  \\ 
        FlowMonitor     & 40.9 & 31.1  & 41.9 & 4.5 & 62.2 & 93.2  \\
        NAT         &  8.2  & 38.6  & 49.1  & 6.4  & 42.1 & 80.7  \\ 
        \bottomrule
    \end{tabular*}
    \captionof{table}{Prediction accuracy comparison under both multi-resource contention and varying traffic attributes. On average \sys reduces MAPE by 78.8\% compared to \slomo, at 3.7\% and 17.5\%, respectively. Unless otherwise stated, rows of table are sorted in ascending order of \sys's MAPE.}
    \vspace{-5mm}
    \label{tab:err_real}
\end{table}
\setlength{\tabcolsep}{6pt}

\subsection{Deep-Dive: Multi-Resource Contention} 
\label{eva:acc_multi}

\begin{figure}[t]
    \centering
    \small
    \includegraphics[width=0.88\linewidth]{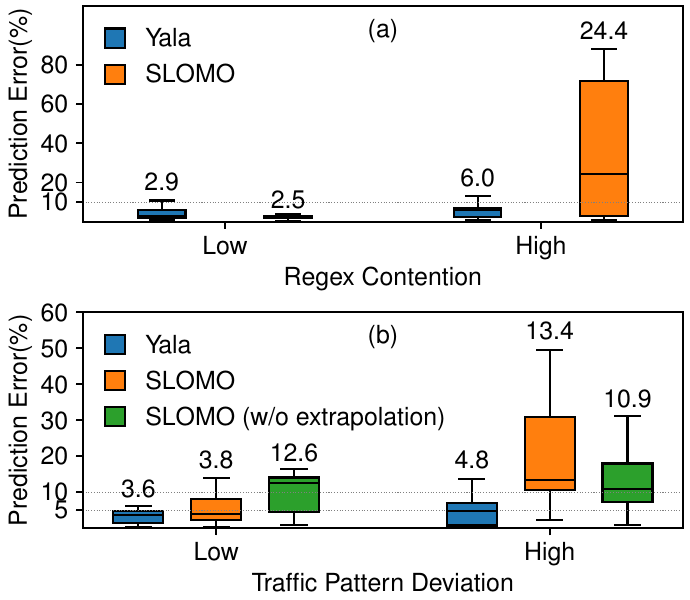}
    \caption{The distribution (box and whisker plot) of the absolute percentage errors of (a) multi-resource contention with different regex contention levels, and (b) memory-only contention with different ranges of variations in the flow count. We show the value of median error in both figures.}
    \Description[]{}
    \vspace{-2mm}
    \label{fig:err_breakdown}
\end{figure}

First we look at how \sys's per-resource modeling and composition approach handles multi-resource contention. 
We fix the traffic profile to be the default in this section to isolate its impact. 
We choose FlowMonitor and NIDS that utilize multiple resources, and co-run each with \membench and \regexbench with varying contention levels.
Table~\ref{tab:error_multi_fixed} presents the comparison.
We can observe that \sys reduces the MAPE by 44.2\% and 17.1\% for FlowMonitor and NIDS, respectively, 
To better analyze the source of accuracy gains, 
we zoom in on FlowMonitor and vary the regex contention level it receives in two ranges:
low range (MTBR $\leq$ 600 matches/MB), and high range (MTBR $>$ 600 matches/MB).
Figure~\ref{fig:err_breakdown}(a) depicts the distribution of the absolute percentage errors.  
\sys maintains low errors as contention rises with median errors consistently below 6.0\%.
With low contention, \slomo also achieves high accuracy with median error at 2.5\%, because in this case multi-resource contention effectively reduces to single-resource (memory for FlowMonitor) contention. 
{Here \sys's error is actually slightly higher that \slomo, because \slomo enjoys the same amount of training data as \sys but concentrate on one fixed traffic profile.
}
During high contention, \slomo's inability to model contention on these two resources simultaneously lead to high median errors at 24.4\%.

\setlength{\tabcolsep}{3pt}
\begin{table}[t]
    \centering
    \footnotesize
    \begin{tabular*}{1\linewidth}{@{\extracolsep{\fill}}ccccccc}
        \toprule
        \multirow{2}{*}{NF} & \multicolumn{3}{c}{SLOMO} & \multicolumn{3}{c}{\sys} \\ 
        \cmidrule(lr){2-4}\cmidrule(lr){5-7}
          & \makecell{MAPE\\(\%)} & \makecell{$\pm$5\% \\ Acc.(\%)} & \makecell{$\pm$10\% \\ Acc.(\%)} &\makecell{MAPE\\(\%)}& \makecell{$\pm$5\% \\ Acc.(\%)} & \makecell{$\pm$10\% \\ Acc.(\%)} \\ 
          \midrule
        NIDS        &  21.4 & 60.0 & 71.2 & 4.3  & 55.6 & 95.6   \\ 
        FlowMonitor     & 49.3  & 18.5  & 33.3 & 5.1 & 59.3 & 88.9 \\
        \bottomrule
    \end{tabular*}
    \caption{Prediction accuracy of SLOMO and \sys when the NF runs under only multi-resource contention. Traffic profile is fixed to the default one.}
    \vspace{-5mm}
    \label{tab:error_multi_fixed}
\end{table}
\setlength{\tabcolsep}{6pt}

To further evaluate the use of \sys's composition design based on execution pattern (\cref{design:resource_compose}),  we write two synthetic NFs, NF1 and NF2, with NF1 using memory and regex, while NF2 adds the compression accelerator. 
Each has both a pipeline and run-to-completion version. 
We compare \sys to the simple sum and min composition (recall \cref{moti:multi-resource}), where they use the same per-resource models trained specifically for NF1 and NF2.
As shown in Table~\ref{tab:compose}, \sys attains the best accuracy across all cases and resource usage patterns, with MAPE lower than 2\%. 
This gain essentially comes from properly modeling NF's execution pattern to capture the end-to-end impact of multi-resource contention.

\begin{table}[t]
    \centering
    \small
    \begin{tabular*}{1\linewidth}{@{\extracolsep{\fill}}ccccc}
    \toprule
    \multirow{2}{*}{NF}  & \multirow{2}{*}{Pattern}  & \multicolumn{3}{c}{MAPE (\%)}  \\ 
    \cmidrule(lr){3-5}
      &   & sum & min & \sys  \\ 
    \midrule
    \multirow{2}{*}{NF1} & pipeline & 9.8 & 0.7 & 0.7\\
    & run-to-completion & 8.7 &  12.4 & 1.3 \\
    \multirow{2}{*}{NF2} & pipeline & 21.9 & 1.8 & 1.8 \\
    & run-to-completion & 14.6 & 7.6 & 1.6 \\
    \bottomrule
    \end{tabular*}
\captionof{table}{Prediction error of different multi-resource composition approaches for different execution patterns.}
\vspace{-8mm}
\label{tab:compose}
\end{table}

\subsection{Deep-Dive: Traffic Attributes} 
\label{eva:acc_traffic}
We now move to examine \sys's design on traffic-aware modeling.
We choose traffic-sensitive NFs and co-run each with \membench on a single \bftwo.
Since \slomo only considers memory contention, we set a fixed contention level in memory, exclude other resource contention, and generate 100 traffic profiles by randomly changing number of flows, packet size, and MTBR when applicable for each NF.
Table~\ref{tab:error_single_dynamic} shows the results aggregated from all profiles.
Again \sys attains superior accuracy over \slomo across the board, with over 90\% in \textit{$\pm10\%$ Acc.} and $<$5\% MAPE for most NFs.

We then zoom into one particular attribute, flow count, which \slomo also specifically models. 
We vary the flow count between training and testing across two ranges: low range where it changes by at most 20\%, and high range ($>$20\%).
Figure~\ref{fig:err_breakdown}(b) displays the distribution of absolute percentage error in this case.
\sys maintains low errors consistently, with a median at 4.7\%, highlighting the benefits of our traffic-aware modeling (\cref{sec:tfaware}).
Under low-range variations, \slomo exhibits high accuracy with sensitivity extrapolation~\cite{slomo20}. 
However, as traffic profiles undergo more significant changes, 
\slomo suffers from high prediction errors, with 13.4\% at the median. This is consistent with~\cite{slomo20}: the extrapolation only works when the NF's sensitivity profile in training has enough overlap with that under testing traffic~\cite{slomo20}, which corresponds to low range profiles here.

\setlength{\tabcolsep}{3pt}
\begin{table}[t]
    \centering
    \footnotesize
    \begin{tabular*}{1\linewidth}{@{\extracolsep{\fill}}ccccccc}
        \toprule
        \multirow{2}{*}{NF} & \multicolumn{3}{c}{SLOMO} & \multicolumn{3}{c}{\sys} \\ 
        \cmidrule(lr){2-4}\cmidrule(lr){5-7}
          & \makecell{MAPE\\(\%)} & \makecell{$\pm$5\% \\ Acc.(\%)} & \makecell{$\pm$10\% \\ Acc.(\%)} &\makecell{MAPE\\(\%)}& \makecell{$\pm$5\% \\ Acc.(\%)} & \makecell{$\pm$10\% \\ Acc.(\%)} \\ 
          \midrule
        NIDS        & 2.5 & 90.0 &  100.0 & 1.1 & 98.0 & 100.0  \\ 
        FlowClassifier  & 10.4 &  52.0 & 66.0 & 2.9 & 80.0 & 100.0 \\
        NAT         & 9.5  & 28.0  & 64.0 & 3.1 & 82.0 & 96.0 \\ 
        FlowTracker   & 4.0 &  70.0 & 92.0 & 3.5   & 74.0 & 96.0 \\
        FlowStats   & 9.5 & 44.0 & 66.0 & 4.7  &  72.0 & 92.0  \\ 
        FlowMonitor     & 11.9 &  20.0 & 44.0 & 4.8   & 62.0 & 88.0 \\ 
        IPTunnel    & 88.0 & 24.0  & 52.0  & 5.6 & 80.0 & 94.0 \\ 
        \bottomrule
    \end{tabular*}
    \captionof{table}{Prediction accuracy of SLOMO and \sys when target NF runs under memory-only contention and dynamic traffic profiles.}
    \vspace{-5mm}
    \label{tab:error_single_dynamic}
\end{table}
\setlength{\tabcolsep}{6pt}

\subsection{\sys Use Cases}
\label{eva:usecase}

We now illustrate \sys's practical benefits through two use cases: (1) It enables contention-aware scheduling of NFs to improve resource utilization; (2) it facilitates performance diagnosis for NFs with dynamic traffic.

\subsubsection{Contention-Aware Scheduling}
\label{eva:schedule}
We consider the scenario in which the operator places the NFs as they arrive to a group of \snics to maximize resource utilization (\ie minimize \snics used) while maintaining their SLAs. 
SLA here is defined as the maximum allowed throughput drop relative to the baseline when the NF runs solo.
Given that the offline version of this problem is NP-complete bin-packing~\cite{binpack,binpack1}, we follow previous works~\cite{slomo20,resq18,octans19} and consider online heuristics that deploy the NFs one by one.  
Specifically, we compare the following strategies: 
(1) Monopolization, which forbids co-location of NFs; (2) Greedy, which places an NF onto the \snic with most available resources~\cite{e319,meili23}; (3) Contention-aware, which first predicts the performance of all NFs on a \snic if the current NF gets deployed onto this \snic, and then deploys the NF onto the \snic if no SLA violation is predicted. 
Both \sys and \slomo can provide such contention-aware predictions. A new \snic is added to the cluster when there is no feasible placement.

Table~\ref{tab:overhead} compares the above strategies over {100 random sequences of 500 NF arrivals each.} 
Each NF is assigned the default traffic profile, and its SLA is set to 5-20\% throughput drop.
We examine resource wastage, \ie how many additional \nics are used against the optimal plan found by exhaustive search, and the corresponding SLA violations.
We observe that \sys minimizes resource wastage to merely 0.5\% and reduces SLA violations by 88.5\% and 92.2\% over Greedy and \slomo on average. 
The near-optimal performance of \sys illustrates its potential in coordinating NF scheduling in a real-world scenario. 

\begin{table}[t]
    \centering
    \small
    \begin{tabular*}{1\linewidth}{@{\extracolsep{\fill}}ccc}
    \toprule
    Approach & Resource Wastage (\%) & SLA Violations (\%)  \\ 
    \midrule
    Monopolization & 196.3 & 0 \\
    Greedy & 19.0 &  16.5 \\
    SLOMO & -21.8 &  24.4 \\
    \sys & 0.5 &  1.9 \\
    \bottomrule
    \end{tabular*}
\captionof{table}{
    \sys's usecase in contention-aware scheduling compared to other baseline strategies. 
    \slomo's negative resource overhead stems from erroneous placements compared to the optimal deployment.}
\vspace{-2mm}
\label{tab:overhead}
\end{table}

\begin{table}[t]
    \centering
    \small
    \begin{tabular*}{1\linewidth}{@{\extracolsep{\fill}}ccc}
    \toprule
    \multirow{2}{*}{NF}  & \multicolumn{2}{c}{Correctness (\%)} \\
     \cmidrule(lr){2-3}
      & \slomo & \sys  \\ 
    \midrule
    Flowstats & 100.0 & 100.0 \\
    FlowMonitor & 38.7 & 100.0 \\
    IPComp Gateway & 29.3 & 100.0  \\
    \bottomrule
    \end{tabular*}
\captionof{table}{Percentage of correct identifications of performance bottleneck using \slomo and \sys.}
\vspace{-5mm}
\label{tab:diagnosis}
\end{table}

\subsubsection{Performance Diagnosis}
\label{eva:diagnosis}
In practice, performance diagnosis has important values as it allows programmers to systematically explore the design spaces, identify performance bottlenecks and optimization opportunities, and even provide early-stage insights/guidances on next-generation \snic~\cite{morpheus22,slomo20}. 
Here we show another usecase of \sys in diagnosing performance bottlenecks in NFs with dynamic traffic, when the bottleneck may shift across resources.

As Table~\ref{tab:diagnosis} shows we deploy FlowStats, FlowMonitor, and IPComp Gateway that all use regex accelerator.
We co-run each of them with \membench and \regexbench and adjust the MTBR from 0 to 1100 matches/MB while keeping memory contention levels unchanged, and manually analyze the actual its performance bottleneck using the hotspot analysis function of \texttt{perf-tools}~\cite{perf}.
This is the ground truth. 
We then calculate the percentage of correct identification of bottleneck using our prediction models. 
We can see \sys accurately identifies bottleneck for all three NFs with its multi-resource performance modeling, while \slomo only works for FlowStats, as it is always bottlenecked on memory. 
FlowMonitor and IPComp Gateway's bottleneck actually shifts with traffic.
For example, we observe that FlowMonitor's bottleneck is memory with MTBR at 80 matches/MB, but changes to regex with MTBR at 1000 matches/MB (default traffic profile).
Thus this usecase demonstrates \sys's capability in pinpointing NF bottlenecks with dynamic traffic.

\subsection{Adaptive Profiling}
\label{eva:micro}
\begin{figure}[t]
    \centering
    \small
    \includegraphics[width=0.88\linewidth]{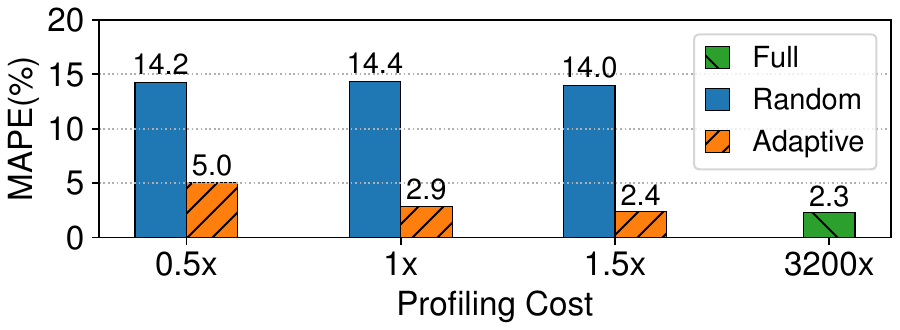}
    \caption{Prediction error on FlowClassfier using full, random and \sys's adaptive profiling. We change the profiling quota of random and adaptive profiling to 0.5$\times$ and 1.5$\times$ of that used in Table~\ref{tab:adaptive_profile}.}
    \Description[]{}
    \label{fig:adaptive_detail}
\end{figure}

Lastly we examine the adaptive profiling design of \sys. 
We select traffic-sensitive NFs from Table~\ref{tab:nfs}, and train them using three different approaches --- full profiling, random profiling, and adaptive profiling (\cref{design:adaptive}).
For random profiling and \sys's adaptive profiling, we set the same number of training data points (\aka profiling quota) to ensure a fair comparison. 
For full profiling, we use 80\% of the profiled data for training, and the remaining 20\% for testing.
Profiling cost is represented by the number of training data samples  normalized against adaptive/random profiling's quota. 
We observe from Table~\ref{tab:adaptive_profile} that \sys's adaptive profiling offers comparable accuracy to full profiling which uses 3200$\times$ more data\footnote{This is because we use 16 values for packet sizes and 200 values for the number of flows in full profiling. For each (packet size, number of flows) tuple, full profiling repeats profiling over a set of random memory contention levels.}. 
Compared to random, adaptive profiling significantly enhances accuracy within the same profiling quota. 
We further show the benefit of adaptive profiling using FlowClassifier as an example in Figure~\ref{fig:adaptive_detail}.
We adopt the same setting as Table~\ref{tab:adaptive_profile} but change the profiling quota to 0.5$\times$ and 1.5$\times$.
Figure~\ref{fig:adaptive_detail} reveals that by increasing current profiling quota by 50\% (still $\sim$2100$\times$ less than the profiling cost of full profiling), adaptive profiling achieve similar error compared to full profiling, at 2.4\% and 2.3\%, respectively, while random profiling does not exhibit accuracy improvement since performance-significant ranges of traffic attribute is still not covered in the training data. 

\setlength{\tabcolsep}{3pt}
\begin{table}[t]
    \centering
    \footnotesize
    \begin{tabular*}{1\linewidth}{@{\extracolsep{\fill}}ccccccc}
        \toprule
        \multirow{3}{*}{NF} & \multicolumn{2}{c}{Full} & \multicolumn{2}{c}{Random} & \multicolumn{2}{c}{Adaptive} \\ 
        \cmidrule(lr){2-3} \cmidrule(lr){4-5} \cmidrule(lr){6-7} 
        & \multicolumn{2}{c}{\textbf{P.C.}: 3200$\times$} & \multicolumn{2}{c}{\textbf{P.C.}: 1$\times$} &  \multicolumn{2}{c}{\textbf{P.C.}: 1$\times$} \\
        \cmidrule(lr){2-3} \cmidrule(lr){4-5} \cmidrule(lr){6-7}
           & MAPE & $\pm$10\%&  MAPE& $\pm$10\% &  MAPE  & $\pm$10\%  \\ 
           & (\%) & Acc.(\%)  &  (\%) & Acc.(\%)  & Acc.(\%) & (\%) \\ 
           \midrule
        FlowClassifier  & 2.3 & 100.0 &  14.4 & 28.0  & 2.9 & 100.0  \\
        NAT         & 2.9  & 98.0  &   9.6 & 62.0 &  3.1 & 96.0   \\ 
        FlowTracker   & 3.4 & 96.0  &   38.9 & 0.0 &  3.4 & 86.0 \\
        FlowMonitor     & 4.5  & 86.0 &   12.3 & 42.0 &  4.8 & 88.0 \\ 
        FlowStats   & 5.3 & 90.0 &   9.8 & 68.0 &   4.9 & 90.0   \\ 
        IPTunnel    & 5.3 & 98.0 &   8.5 & 82.0 &  5.9 & 96.0 \\ 
        \bottomrule
    \end{tabular*}
\captionof{table}{Profiling cost and model accuracy using full, random and \sys's adaptive profiling. \textbf{P.C.} refers to profiling cost.}
\label{tab:adaptive_profile}
\end{table}
\setlength{\tabcolsep}{6pt}

\noindent\textbf{Time cost of profiling.}
\label{eva:overhead}
As discussed in \cref{sec:models}, \sys requires offline profiling for each NF to build per-resource models and identify the execution pattern. 
This process primarily involves the collection of: (1) the contention level of synthetic benchmark NFs (\membench, etc.), and (2) the contention level and sensitivity profiles of the target NF. 
Across our experiments, we find that on average 1.6 hours and 0.5 hours for each, respectively. 
These time investments are acceptable since profiling is a one-time effort.

\section{Discussion}
\label{sec:discuss}
We discuss a few issues one may have regarding \sys.

\noindent{\textbf{What if the configuration of an NF changes?}} 
It is possible that an NF's configuration is adjusted for various reasons, e.g. adjusting ACL rules in a Firewall NF to apply new policies.
Such changes can cause performance characteristics to change, making the existing model inaccurate. 
To make \sys ``configuration-aware'', one may adopt a similar approach of traffic attributes, \ie extracting ``configuration attributes'' for an NF and integrating it into the per-resource models.
We leave this as future work.

\noindent{\textbf{Can \sys be generalized to other \snics?}}
\sys can be generalized to other SoC \snics due to the similar architecture of their hardware accelerators and memory subsystem.
To quickly validate such generalizability, we collect data and train performance models of a Firewall NF~\cite{db23} running on an AMD Pensando
SmartNIC~\cite{pensando}.
The NF conducts a flow walk on hardware flow table and updates entry metadata upon matching against flows in the input traffic.
As shown in Table~\ref{tab:psd}, the average prediction error of \sys is 0.9\%, which is 17.5\% lower compared to that of \slomo.
This result reflects that applying \sys to other SoC \snics is feasible.
However, for \snics like on-path \snics whose architecture significantly deviates from SoC \snics, more investigation is needed into the contention behavior of their specialized hardware resources~\cite{switchvm24,menshen22}.

\setlength{\tabcolsep}{3pt}
\begin{table}[t]
    \centering
    \footnotesize
    \begin{tabular}{ccccccc}
        \toprule
        \multirow{2}{*}{NF} & \multicolumn{3}{c}{SLOMO} & \multicolumn{3}{c}{\sys} \\ 
        \cmidrule(lr){2-4}\cmidrule(lr){5-7}
          & \makecell{MAPE\\(\%)} & \makecell{$\pm$5\% \\ Acc.(\%)} & \makecell{$\pm$10\% \\ Acc.(\%)} &\makecell{MAPE\\(\%)}& \makecell{$\pm$5\% \\ Acc.(\%)} & \makecell{$\pm$10\% \\ Acc.(\%)} \\ 
          \midrule
        Firewall        &  18.4 & 58.7 & 64.7 & 0.9  & 100.0 & 100.0   \\ 
        \bottomrule
    \end{tabular}
    \caption{Prediction accuracy of SLOMO and \sys when target NF runs under memory-only contention and dynamic traffic profiles on Pensando \snic.}
    \vspace{-5mm}
    \label{tab:psd}
\end{table}
\setlength{\tabcolsep}{6pt}

\noindent{\textbf{Can \sys adapt to system-wide state changes that affect the maximum performance of NFs?}} 
\sys predicts the maximum throughput of co-located NFs to ensure they conform to SLA under multi-tenancy. 
However, \snics may experience system-wide state changes due to various reasons, altering the maximum throughput of NFs. 
A typical example is dynamic voltage and frequency scaling (DVFS)~\cite{dvfsarm,dvfs10}, which may affect the NF's maximum performance due to CPU frequency down-scale or up-scale.
Different DVFS policies can be enabled by setting the frequency scaling governor of CPUs on servers, which is however not supported on common SoC \snics currently, e.g. the \snics we use in our experiments (Nvidia Bluefield-2 and AMD Pensando)~\cite{bfscripts}. 
If we are to extend \sys to predict maximum performance of on-NIC NFs on future generations of \snics that support DVFS and other similar system-wide state changes, related variables, e.g. OS policy and thermal state for DVFS, should be integrated in our models to ensure high accuracy under different system states, e.g. power state in DVFS.

\section{Related Work}
\label{sec:related}
\noindent{\bf{NF performance modeling.}} There have been extensive efforts on modeling NF performance.
We compare \sys with past frameworks in Table~\ref{tab:related}.
\sys is, to our knowledge, the first contention-aware framework that explicitly models multi-resource contention and traffic attributes.

\setlength{\tabcolsep}{4pt}
\begin{table}[t]
    \centering
    \small
    \begin{tabular*}{1\linewidth}{@{\extracolsep{\fill}}cccccc}
        \toprule
        Framework  & \makecell{Contention-\\aware} & \makecell{Multi-\\resource} & \makecell{Traffic-\\aware} & \makecell{Sourcecode-\\agnostic} \\ 
        \midrule
        
        Clara~\cite{clara21}    &  \tikzxmark & \tikzcmark & \tikzcmark & \tikzxmark \\ 
        LogNIC~\cite{lognic23}    &  \tikzxmark  & \tikzcmark  &  \tikzcmark  & \tikzcmark  \\
        BubbleUp~\cite{bubbleup11}    & \tikzcmark & \tikzxmark & \tikzxmark & \tikzcmark \\
        \slomo~\cite{slomo20}    & \tikzcmark & \tikzxmark & \tikzhalfcmark & \tikzcmark \\
        \textbf{\sys}    &  \tikzcmarkbf  & \tikzcmarkbf  & \tikzcmarkbf  & \tikzcmarkbf   \\ 
        \bottomrule
    \end{tabular*}
    \caption{NF performance prediction frameworks. 
    ``Sourcecode-agnostic'' means the framework does not require NF's source code. \slomo only considers 20\% variation in flow counts so it is considered ``half'' traffic-aware.}
    \label{tab:related}
    \vspace{-5mm}
\end{table}
\setlength{\tabcolsep}{6pt}

\noindent{\bf{Isolation.}}
Resource isolation techniques have been explored to provide performance guarantees of co-running applications~\cite{fairnic20,parties19,resq18,picnic20,netbricks,osmosis24}.
For example, FairNIC~\cite{fairnic20} proposes isolation solutions for \snic accelerators. PARTIES~\cite{parties19} and ResQ~\cite{resq18} leverages off-the-shelf isolation techniques, \eg Intel CAT~\cite{intelcat} to enable QoS-aware resource partitioning.
However, these efforts are either inapplicable to \snics, or provide only partial isolation, or require substantial rewriting of NFs.

\noindent{\bf{\snic-accelerated NFV.}}
NFV platforms have been leveraging \snics to improve energy efficiency and enable host resource-saving~\cite{e319,ipipe19,exoplane23,uno17,db23}. For example, E3~\cite{e319} builds a \snic-accelerated microservice execution platform with high energy efficiency. 
\sys is complementary to them as it can assist operators to make better runtime decisions, thus improving resource utilization and reducing SLA violations.
\section{Conclusion}
\label{sec:conclusion}
Prior contention-aware performance prediction frameworks fail to accurately predict the performance of on-\nic NFs due to multi-resource contention and changing traffic profiles. 
We systematically analyze multi-resource contention characteristics on \snic as well as the impact of traffic attributes on performance of on-\nic NFs. 
Our insights enable the design of \sys, a multi-resource contention-aware and traffic-aware performance prediction framework for on-NIC NFs. 
\sys achieves accurate performance predictions, with 3.7\% prediction error and 78.8\% accuracy improvement on average compared to prior works, and enables new usecases.
\appendix
\section{Artifact Appendix}

\subsection{Abstract}
This artifact contains source code and related tools for \sys, a multi-resource contention- and traffic-aware performance prediction framework for on-NIC NFs.
\sys is publicly available on Github~\cite{yalagithub}. Specifically, we provide source code of model training and prediction.
Example profiles of an NF (FlowMonitor) can be used to train its models and produce predictions on its throughput for demonstration.  
In addition, we open-source the benchmark NFs (\membench, \regexbench and \comprebench), real NFs, NF framework (a modified version of Click), rulesets and related tools mentioned in our paper.

\subsection{Artifact check-list (meta-information)}
{\small
\begin{itemize}
    \item {\textbf{Compilation:} We provide python scripts for training and prediction, which do not require a compiler. Compiling \texttt{Click} and NFs requires \texttt{gcc}~\cite{click2000}. Compiling rulesets requires \texttt{RXP compiler}.} \texttt{Makefiles} are provided for compilation.
    \item {\textbf{Binary:} We provide python scripts for training and prediction instead of binaries. Some binaries of related tools are included, \eg compiled ruleset.}
  \item {\textbf{Model:} Gradient boosting regression and linear regression from \texttt{sklearn}~\cite{sklearn}.}
  \item {\textbf{Hardware:} We use NVIDIA BlueField-2 MBF2H332A-AENOT SmartNIC~\cite{bf21}. The performance counters in Table~\ref{tab:pcs} can be accessed on this \snic.}
  \item {\textbf{Publicly available?:} Yes, on \url{https://github.com/NetX-lab/Yala}}
  \item {\textbf{Code licenses?:} BSD 3-Clause "New" or "Revised" License}.
  \item {\textbf{Archived?:} Yes, on \url{https://doi.org/10.5281/zenodo.14051092}}
\end{itemize}
}
    \begin{table}[ht]
    \centering
    \small
    \begin{tabular}{ll}
        \toprule
        Counter &  Definition \\ 
        \midrule
        IPC  & Instructions per cycle.      \\ 
        IRT  & Instruction retired.        \\ 
        L2CRD  & L2 data cache read access.  \\ 
        L2CWR &  L2 data cache write access.  \\ 
        MEMRD &  Data memory read access.       \\ 
        MEMWR & Data memory write access. \\
        WSS & Working set size. \\
        \bottomrule
    \end{tabular}
    \caption{\small{Performance counters for training the per-resource model of memory subsystem.}}
    \vspace{-3mm}
    \label{tab:pcs}
\end{table}

\subsection{Description}

\subsubsection{How to access}
Yala is publicly available on Github: \url{https://github.com/NetX-lab/Yala}. 

\subsubsection{Hardware dependencies}
We profile NFs on NVIDIA BlueField-2 MBF2H332A-AENOT SmartNIC. 
The traffic generator uses another ConnectX-6 100GbE NIC.
The training and prediction do not have specific hardware requirements.
\subsubsection{Software dependencies}
Dependencies of the software are listed as below:
\begin{itemize}
\item Training and prediction
    \begin{itemize}
        \item \texttt{Python}: 3.8
        \item \texttt{scikit-learn}~\cite{sklearn}: 0.24.2
        \item \texttt{numpy}: 1.19.5
        \item \texttt{pandas}: 1.1.5
        \item \texttt{tabulate}: 0.9.0
    \end{itemize}
    \item Traffic generator
    \begin{itemize}
        \item \texttt{DPDK-Pktgen}~\cite{pktgen}: 23.03.1
    \end{itemize}
    \item NF frameworks
    \begin{itemize}
        \item \texttt{Click}~\cite{click2000}: 2.1
        \item \texttt{DPDK}~\cite{DPDK}: MLNX\_DPDK\_20.11.6
        \item \texttt{DOCA}\cite{doca1}: 1.5.0-LTS
    \end{itemize}
\end{itemize}

\subsection{Installation and Testing}
For installation of hardware and software dependencies, please follow our official guide on Github. 
The training and prediction of \sys can be tested upon satisfying software dependencies.

We provide an example of training and using the model to predict throughput for FlowMonitor.
To train \sys using example training set, please go to \texttt{/model} directory and run \texttt{python3 train.py} command. This generates a \texttt{models.pkl} file containing linear model, memory-only model (\slomo), regex-only model and \sys's model for FlowMonitor as an example.
Then to evaluate the model accuracy on the example test set, run \texttt{python3 predict.py} command.
This reports MAPE, \textit{$\pm5\%$ Acc.} and \textit{$\pm10\%$ Acc} of the four models mentioned.

\subsection{Other Notes}
Detailed instructions on using open-sourced tools, \eg benchmark NFs, can be found in "additional tips" on our Github.

\subsection{Methodology}

Submission, reviewing and badging methodology:

\begin{itemize}
  \item \url{https://www.acm.org/publications/policies/artifact-review-and-badging-current}
  \item \url{https://cTuning.org/ae}
\end{itemize}

\clearpage
\balance
\bibliographystyle{plain}
\balance
\bibliography{main}
\clearpage
\end{document}